\documentclass{elsart1p}
\usepackage{graphicx}
\usepackage{amssymb}

\newcommand{\mev}{\textrm{ MeV}}
\newcommand{\be}{\begin{equation}}
\newcommand{\ee}{\end{equation}}
\newcommand{\ba}{\begin{eqnarray}}
\newcommand{\ea}{\end{eqnarray}}

\begin{document}
\begin{frontmatter}
% use the corauthref command within \author for corresponding author
% footnotes;
\title{Meson and Baryon resonances}
\author{E. Oset$^1$, L.S. Geng$^1$, D. Gamermann$^1$, R. Molina$^1$, D.
Nicmorus$^1$,
 }
\author{ J. Yamagata-Sekihara$^2$,  H. Nagahiro$^2$, S. Hirenzaki$^2$,}
\author{  D. Jido$^3$, M. D\"oring$^4$ and A. Ramos$^5$}

\address{$^1$ Departamento de Fisica Teorica and IFIC, Universidad de Valencia,
Spain}
\address{$^2$ Department of Physics, Nara Women University, Japan}
\address{$^3$ Yukawa Institute for Theoretical Physics, Kyoto University, Japan}
\address{$^4$ IKP, J\"uelich,Germany}
\address{$^5$ Departament d'Estructura i Constituents de la Materia, Universitat
de Barcelona, Spain}
\begin{abstract}
In this talk I review recent advances on the structure of the meson and baryon
resonances which can be dynamically generated from the interaction of mesons
or mesons and baryons. Particular emphasis is put on results involving vector
 mesons, which
bring new light into the nature of some of the observed higher mass
mesons and baryons and make predictions for new states. 
% Text of abstract
\end{abstract}
\end{frontmatter}
%
% main text
\section{Introduction}
  The combination of chiral Lagrangians and unitarity in coupled channels
has given rise to a new area of research, $U\chi PT$, which has proved very rich,
enlarging the realm of applicability of the information contained in the
chiral Lagrangians much beyond what the perturbative scheme of $\chi PT$ allows.
The nonperturbative scheme, unitarized chiral perturbation theory ( $U\chi PT$ ) 
provides a
reliable method to study meson baryon interactions and gives rise to a wealth of
dynamically generated meson and baryon states, resulting from the
interaction, which do not qualify as standard $q \bar{q}$ or $3q$ states
\cite{review}. Typical examples are the generation of low lying scalar 
meson states from the interaction of two pseudoscalars, the generation of low
lying axial vector mesons from the interaction of a pseudoscalar and a vector, 
the generation of
$J^P=1/2^-$ baryons from the interaction of one pseudoscalar and one baryon of
the octet or  $J^P=3/2^-$ baryons from the interaction of a pseudoscalar and one
baryon of the decuplet. 

  Very recently the interaction of vector mesons among themselves or with
  baryons has been tackled and novel results are appearing that will be reported
  here.  The scheme that makes this study possible is the hidden gauge approach
  for vector mesons, pseudoscalars and photons \cite{hidden2}, 
which contains the basic chiral Lagrangians for the interaction of
pseudoscalars and extends the scheme to include explicitly vector mesons with
the interaction among themselves and with other hadrons. We expose the basic
ideas in the following section.

\section{The hidden gauge formalism}

The HGS formalism to deal with vector mesons
\cite{hidden2} is a useful and internally
consistent scheme which preserves chiral symmetry. In this
formalism the vector meson fields are gauge bosons of a hidden local
symmetry transforming inhomogeneously. After taking the unitary 
gauge, the vector meson fields transform exactly in the manner as 
in the non linear realization of chiral symmetry.
The Lagrangian involving pseudoscalar mesons, photons and
vector mesons can be written as

\ba
{\cal L}={\cal L}^{(2)}+ {\cal L}_{III}; ~~~{\cal L}^{(2)}=\frac{1}{4}f^2\langle D_\mu U D^\mu U^\dagger+
 \chi U^\dagger+\chi^\dagger U\rangle \\
 {\cal L}_{III}=-\frac{1}{4}\langle V_{\mu\nu}V^{\mu\nu}\rangle
 +\frac{1}{2}M_V^2\langle [V_\mu-\frac{i}{g}\Gamma_\mu]^2\rangle , 
 \label{eq:LIII}
 \ea
where $\langle ...\rangle$ represents a trace over $SU(3)$ matrices. The
covariant derivative is defined by
\be
D_\mu U=\partial_\mu U-ieQA_\mu U+ieUQA_\mu,
\ee
with $Q=diag(2,-1,-1)/3$, $e=-|e|$ the electron charge, and $A_\mu$
the photon field.
The chiral matrix $U$ is given by $U=e^{i\sqrt{2}\phi/f}$
with $f$ the pion decay constant ($f=93\mev$). The $\phi$ and
$V_\mu$ matrices are
the usual $SU(3)$ matrices containing the pseudoscalar mesons and
vector mesons respectively.

In this formalism one finds cancellations among terms, such that
 ultimately the photon
couples to the pseudoscalars or vector mesons through direct coupling to a
neutral vector, the basic feature of vector meson dominance, VMD.

In  ${\cal L}_{III}$, $V_{\mu\nu}$ and $\Gamma_\mu$ are  defined as 
\be V_{\mu\nu}=\partial_\mu
V_\nu-\partial_\nu V_\mu-ig[V_\mu,V_\nu];~\Gamma_\mu=\frac{1}{2}[u^\dagger(\partial_\mu-ieQA_\mu)u
+u(\partial_\mu-ieQA_\mu)u^\dagger]
\ee
with $u^2=U$. The hidden gauge coupling constant $g$ is related to $f$ and the
vector meson mass ($M_V$)
through $g=\frac{M_V}{2f}$, 
which is one of the forms of the KSFR relation.

With these lagrangians one can then study meson-meson 
(pseudoscalars), meson-vector, meson-baryon, vector-vector or vector-baryon
interactions. For the vector-vector case one has the contact term of the four
vectors Lagrangian or the 
exchange of vector mesons using the three vector vertices. 
From the $\langle V_{\mu\nu} V^{\mu\nu}\rangle$ term
of ${\cal L}_{III}$, (see Eq.~(\ref{eq:LIII})), one 
obtains the coupling of three vector mesons and four vector mesons, which are
  essential in the present work.\\

{\it 2.1 Axial vector mesons.}

  This is one example where this theory shows its value. The $\it t$-exchange of a
  vector meson between a vector and a pseudoscalar leads to this interaction,
  which in the limit of neglecting the three momentum of the particles 
  versus the mass of the vector mesons gives the effective chiral Lagrangian
  used in \cite{lutzaxial,luisaxial} by means of which the low
  lying axial vector resonances, $a_1, b_1, K_1, f_1,h_1$ are dynamically 
generated.  An interesting novelty to be recalled in this presentation is the
fact that in \cite{luisaxial} two states for the $K_1(1270)$ were found while
the $K_1(1400)$ did not show up in the approach as a pole of the t-matrix. 
The novel interesting thing to note is that there was already experimental
information supporting the existence of the two $K_1(1270)$ states in the 
$K^- p \to K^- \pi^+  \pi^-  p$ reaction of \cite{daum}, however the
experimental analysis had been forced with just one resonance at the expense of
introducing a large background, much larger than the resonance contribution,
which upon interference produced the  experimental shape of the $K^* \pi$ and
$\rho K$ mass distributions. However, in a reanalysis of the
reaction \cite{gengmurcia} it was shown that two poles can be found at 
(1195-i123) MeV and (1284-i73) (the width is double the imaginary part), the first
one coupling mostly to  $K^* \pi$ and the second one to $\rho K$. According to
this, an experiment where the resonance is excited and leads to $K^* \pi$ in
the final state should put more weight on the first resonance and if it
produces $\rho K$ on the second resonance, as a consequence of which, the peaks
of these two reactions should be displaced by about 90 MeV and the width in the
second experiment should be narrower. This is indeed what the experiment of
\cite{daum} shows and it was naturally interpreted in favor of the two
resonances in the reanalysis of \cite{gengmurcia}.\\

{\it 2.2 Scalar and axial vector mesons with charm.}

  There is a special session devoted to charm, so here I will be deliberately
short, just to recall that within this framework several states are dynamically
generated like the scalars $D_{s0}(2317)$ or $D_{0}(2400)$ 
\cite{lutzd,chiang,daniel}, or the axial vector mesons like the X(3872)
\cite{lutzd,chiangax,danielax}. Actually two states with different C-parity
are predicted in \cite{danielax}. Another interesting finding of \cite{daniel}
is a new scalar state of hidden charm nature (mostly $D \bar{D}$), called X(3700)
for the approximate mass, for which there seems to be  experimental
support in the reaction $e^+ e^- \to J/\psi D \bar{D}$ close to threshold of 
\cite{expddbar}, as shown in the analysis of the experiment of \cite{danielexp}.\\

{\it 2.3  $\rho \rho$ interaction.}

In a recent paper the search for dynamically generated states from the 
vector-vector interaction has been tackled for the first time and the 
results are very
interesting. Taking into account the contact term and the vector exchange terms
in the $\it t$- and $\it u$-channel one finds in \cite{raquel} that this interaction leads
naturally to two states, one scalar, which can be identified with the
$f_0(1370)$ although the mass appears around 1500 MeV, and another state of spin 2,
which can be neatly identified with the $f_2(1270)$. Once one includes the $\pi
\pi$ decay channel, provided also by the theory through a box diagram coming
from the $\rho$ decay into two pions, the agreement of mass and width is
acceptable. It is interesting to note that the theory provides an interaction 
which is stronger for the case of the tensor state, leading naturally to a more
bound state for the tensor than for the scalar. 

   A further support for this association to the physical states comes from the 
study of the radiative decay of these resonances into $\gamma \gamma$
\cite{junko}.  The
evaluation is quite easy since in the hidden gauge approach the photon couples
to the hadrons always through a direct coupling to the neutral vector mesons
(vector meson dominance). Thus the amplitude for this radiative decay is a factor
times the coupling of the resonance to $\rho^0 \rho^0$ in this case, which is
obtained from the residues at the pole of the resonance in the $\rho \rho$
scattering amplitude. The interesting results are 
$\Gamma(f_0(1370)\rightarrow\gamma\gamma)=0.54~~{\rm keV}$, 
$\Gamma(f_2(1270)\rightarrow\gamma\gamma)=2.6~~{\rm keV}$. 
The PDG quotes the
result $\Gamma(f_2(1270) \to \gamma \gamma)= 2.71 ^{+ 0.26} _{-0.23}$ keV
\cite{Amsler:2008zz}.
The situation of the $\gamma \gamma $ decay of the $f_0(1370)$ is rather
unclear. The latest edition of the PDG \cite{Amsler:2008zz} does not quote any
value, superseding old results which  were  ambiguous.  The Belle
collaboration is pursuing work in this direction, with no final result yet, but
strong hints that the $f_0(1370)$ could actually appear at higher energies,
around $1470$ MeV and with a two gamma width about one order of
magnitude smaller than the one of the  $f_2(1270)$ \cite{uehara}.\\

{\it 2.4 Extension to the SU(3) vector-vector interaction.}

  The generalization of the work of \cite{raquel} to the interaction of the
  octet of vector mesons with themselves has been done recently
  \cite{gengvector} and many states are obtained in the coupled channel
  formalism.  A summary of the results obtained and their possible association
  to states of the PDG are shown in the table below\\

\begin{table}  
 \begin{center}
\begin{tabular}{c|c|cc|ccc}\hline\hline
$I^{G}(J^{PC})$&\multicolumn{3}{c|}{Theory} & \multicolumn{3}{c}{PDG data}\\\hline
              & pole position &\multicolumn{2}{c|}{real axis} & name & mass & width  \\
              &               & $\Lambda_b=1.4$ GeV & $\Lambda_b=1.5$ GeV &           \\\hline
$0^+(0^{++})$ & (1512,51) & (1523,257) & (1517,396)& $f_0(1370)$ & 1200$\sim$1500 & 200$\sim$500\\
$0^+(0^{++})$ & (1726,28) & (1721,133) & (1717,151)& $f_0(1710)$ & $1724\pm7$ & $137\pm 8$\\
$0^+(1^{++})$ & (1802,78) & \multicolumn{2}{c|} {(1802,49)}   & $f_1$\\
$0^+(2^{++})$ & (1275,2) & (1276,97) & (1275,111) & $f_2(1270)$ & $1275.1\pm1.2$ & $185.0^{+2.9}_{-2.4}$\\
$0^+(2^{++})$ & (1525,6) & (1525,45) &(1525,51) &$f_2'(1525)$ & $1525\pm5$ & $73^{+6}_{-5}$\\\hline
$1^-(0^{++})$    & (1780,133) & (1777,148) &(1777,172) & $a_0$\\
$1^+(1^{+-})$    & (1679,235) & \multicolumn{2}{c|}{(1703,188)} & &$b_1$ \\
$1^-(2^{++})$    &  (1569,32) & (1567,47) & (1566,51)& &$a_2(1700)??$
\\\hline
$1/2(0^+)$       &  (1643,47) & (1639,139) &(1637,162)&  $K$ \\
$1/2(1^+)$       & (1737,165) &  \multicolumn{2}{c|}{(1743,126)} & $K_1(1650)?$\\
$1/2(2^+)$       &  (1431,1) &(1431,56) & (1431,63) &$K_2^*(1430)$ & $1429\pm 1.4$ & $104\pm4$\\
 \hline\hline
    \end{tabular}
    
\caption{ Predictions of states  and association to known resonances in
the possible cases. In brackets (Mass, Width). $\Lambda_b$  is the range of a
monopole form factor for vector to two pseudoscalars.}  

\end{center}
\end{table}

\section{Baryonic resonances.}
{\it 3.1 Transition form factors.} 

In the baryonic sector we can quote as interesting developments the
evaluation for the first time of transition form factors for resonances
dynamically generated. This has been done for the case of the $N^*(1535)$ in
\cite{jidofactor}, where the $A_{1/2}$ and $S_{1/2}$ amplitudes for protons and
neutrons for the $N^*(1535) \to p(n) \gamma^*$ have been evaluated. 
It is  interesting to compare the values of our $p^*$ and $n^*$
helicity amplitudes $A_{1/2}$ at $Q^2=0$ with those of the PDG.
 We obtain $0.065$ GeV$^{-1/2}$ and $-0.052$ GeV$^{-1/2}$ for
the $p^*$ and $n^*$, respectively, versus the values quoted in the PDG, which
include uncertainties from the compilation of data of several analyses,
$0.090\pm 0.030$ GeV$^{-1/2}$ for the $p^*$ and $-0.046\pm 0.027$ GeV$^{-1/2}$
for the $n^*$. As one can see, the agreement, within uncertainties, is good.
 The agreement is particularly good with the recent MAID2007 analyis 
\cite{Drechsel:2007if} of $A_{1/2}^p=66\pm 7\cdot 10^{-3}\mbox{GeV}^{-1/2}$.

    The main discrepancy of the results of \cite{jidofactor} with the data is in
the $Q^2$ dependence. The theoretical results fall faster than experiment at
large $Q^2$, indicating that there should be a contribution  from some
other components of the resonance, most probably a three quark component, 
even if small.\\

{\it 3.2 Mixture of configurations.}

It is also worth mentioning a work which goes one step forward in the
investigation of the nature of the resonances \cite{hyodo}. The work starts from
the realization that in the generation of the $N^*(1535)$ of \cite{inoue} one
needs different subtraction constants for different channels in the dispersion
relation, unlike the case of the two $\Lambda(1405)$ states \cite{cola}. This
is interpreted in \cite{hyodo} as an indication that there are extra components
in the wave function apart from the meson baryon components. A reanalysis of
the data is done by imposing the same natural subtraction constants in all
channels and then adding a CDD pole that accounts for a genuine component of
the resonance of non meson-baryon nature, such as to get the same results
obtained before
with the different subtraction constants.  The exercise in \cite{hyodo} shows
that such a genuine component is needed for the case of the $N^*(1535)$ but not
for the case of the two $\Lambda(1405)$ states. \\

{\it 3.3 Size of dynamically generated resonances.}

This has been a frequently asked question since for most scattering observables
an answer can be obtained without resorting to describe the spatial
distribution of the meson baryon components.  However, recently there has been a
step forward in this direction in the work of \cite{sekihara} where by coupling
the photon to all components of the resonance in the diagonal transition of 
the resonance to the
same resonance, one can obtain the charge distribution of the state. The
results obtained in \cite{sekihara} are interesting. The lower mass
$\Lambda(1405)$ resonance, around 1390 MeV, has a very small radius reflecting its basic $\pi 
\Sigma$ nature where there are cancellations between the cloud of the  $\pi^+$
and the $\pi^-$. But this is not he case in the higher mass resonance around
1426 MeV, which is basically the I=0 combination of $K^- p, \bar{K}^0 n$,
where the cloud of the   $K^-$ is responsible for a large and negative mean
square radius. \\

{\it 3.4 Baryon resonances from the vector meson-baryon interaction.}

  This field is developing fast.  In \cite{angels} the interaction of the octet
  of vector mesons with the octet of baryons of the proton are studied along the
  lines followed for the study of vectors with vectors discussed in the
  previous section.  The results are very interesting. Many states are obtained
  in this  way, all of them degenerate in spin $J^P=1/2^-,3/2^-$. Most of
  the states obtained can be associated to known resonances like 
  the $N^*(1700) (3/2^-)$, $N^*(2080)(3/2^-)$, $N^*(2090)(1/2^-)$, 
  $\Lambda(1800)(1/2^-)$, $\Lambda(2000)(?^?)$,  
  $\Sigma(1750)(1/2^-)$, 
  $\Sigma(1940)(3/2^-)$, $\Sigma(2000)(1/2^-)$,
   $\Xi(2030)(\geq 5/2^?)$.
  
  Developments with the interactions of vector mesons and the decuplet of
  baryons have also started \cite{pedro} with equally interesting results.

 \section{Conclusions}
 We have reported recent developments along the lines of  chiral
 unitary theory for the study of meson-meson or meson-baryon interaction,
 which show that many more mesonic and baryonic states than expected can 
 qualify as dynamically generated. The introduction of elements of the hidden gauge
 formalism to deal with vector mesons has opened a new door and new states appear
 involving vector mesons as components.  We also reported about further steps
 given to investigate the nature of resonances, as well as to determine
 the distribution of electric charge. Altogether, we have given a perspective 
 of the evolution and recent
interests of the field, for which one can anticipate a fruitful future. \\

 This work is partly supported by DGICYT contract number
FIS2006-03438. We acknowledge the support of the European Community-Research Infrastructure
Integrating Activity
"Study of Strongly Interacting Matter" (acronym HadronPhysics2, Grant Agreement
n. 227431)
under the Seventh Framework Programme of EU.

\label{}
% \appendix
%
% \section{}
% \label{}
%

%
\end{document}